\documentstyle[aps,twocolumn,epsfig]{revtex}

\title{Multi-bit gates for quantum computing}

\author{Xiaoguang Wang, Anders S\o rensen, and Klaus M\o lmer \\
{\small
  Institute of Physics and Astronomy, University of Aarhus}\\
{\small DK-8000 \AA rhus C, Denmark}}
\date{\today}

\begin{document}
\draft

\maketitle

\begin{abstract}
We present a general technique to implement 
products of many qubit operators communicating {\it via} a joint 
harmonic oscillator degree of freedom in a quantum computer. 
By conditional displacements and rotations we can implement Hamiltonians
which are trigonometric functions of qubit operators. With such operators we
can effectively implement higher order gates such as Toffoli
gates and C$^n$-NOT gates, and we show that the entire Grover 
search algorithm can be implemented in a direct way.
\end{abstract}

\bigskip
A quantum computer is a device which is capable of coherently processing
information which is stored in a collection of small quantum systems. Much
attention has been devoted to quantum computers due to
the discovery of algorithms which enable a quantum computer to solve
certain computational problems much faster than any classical computer
\cite{shor,grover}. In a quantum computer an algorithm is represented as a
series of unitary operations, and with a set of so-called universal gates
acting  only on  single 
two-level systems and on pairs of two-level systems, it is possible to
produce any unitary evolution on the Hilbert space of a collection
of two-level systems so that any algorithm can be implemented with these
gates \cite{barenco}.   This theorem couples the
development of quantum computing to the theory of classical
computing where a similar theorem exists, and the complexity of various
computational tasks has been analyzed simply by counting the number of
universal gates required to perform the entire computation. Proposals
for practical implementation of quantum computing deal with practical
issues such as identification of quantum systems which can be addressed
by the experimentalist, but which do not decohere with time, and 
it is a particularly interesting task to find ways to implement
the two-bit gates, acting on the joint state of a pair of two-level systems
(with internal states $|0\rangle$ and $|1\rangle$), 
or qubits. 

Starting with the ion trap proposal by Cirac and Zoller 
\cite{cirac}, a number of proposals for quantum computing exist,
where the individual qubits are coupled to a harmonic oscillator
degree of freedom, and where two-bit gates are implemented
by use of the coupling to such a `data-bus'. 
In the ion trap, the internal electronic or hyperfine
states of the ions are coupled to the collective vibrational degree of freedom
due to the recoil during absorption of laser light; quantum dots
may be localized in an optical cavity and communicate {\it via}
a single mode of the optical field \cite{imamoglu},
and it has been proposed to couple Josephson-junction qubits by 
an LC-oscillator mode in an electrical circuit \cite{schoen}.

In the original ion trap proposal \cite{cirac}, the state of one qubit is 
transferred to the data-bus which is then brought into interaction with the
second qubit of the gate. In this proposal it is essential that the state
of the harmonic oscillator is initially cooled to the ground state. In
order to be able to use an oscillator  
which is not initially in a known state, one can use a 
scheme which only virtually excites it \cite{anders99}, so that the
internal states of the ions are completely disentangled from the unknown
state of the oscillator. It
is even possible to use a scheme which dramatically entangles the
qubits with the oscillator degrees of freedom, to only magically
at the end of the operation remove all entanglement and produce 
an effective coupling of two qubits which is completely independent of the
state of the oscillator \cite{milburn,anders00}.
To produce a unitary time evolution of the form $\exp(i\mu
\hat{A}\hat{B}$), where $\hat{A},\ 
\hat{B}$ commute, these proposals use the simple fact that
\begin{eqnarray} 
\exp(i\lambda_1 x\hat{A})\exp(i\lambda_2 p\hat{B})
\exp(-i&&\lambda_1 x\hat{A})\exp(-i\lambda_2 p\hat{B})\nonumber \\
&&=\exp(-i\lambda_1\lambda_2
\hat{A}\hat{B}).
\label{milburn}
\end{eqnarray}
This property can  be
seen from the Baker-Hausdorf relation since
$\hat{A}$ and $\hat{B}$ commute and the commutator
of the oscillator position $x$ and momentum $p$ is a constant. Application
of (\ref{milburn}) 
requires that one can induce interaction Hamiltonians proportional to
$x\hat{A}$ and $p\hat{B}$. Since these operators can be expressed in
terms of lowering and raising operators 
such couplings can be induced using the resonance condition
associated with excitation and deexcitation of the oscillator together
with the implementation of the internal state operators. If $\hat{A}$
and $\hat{B}$ act on different qubits, we obtain a two-qubit gate.
If they both involve many qubits, we can produce multi-particle
entangled states. It has already been shown that 
if we take $\hat{A}=\hat{B} 
=\hat{J}_y=\sum_l \frac{1}{2}\sigma_{yl}$, Eq.~(\ref{milburn}) leads to an 
effective interaction proportional to $\hat{J}_y^2$, 
which can be used to produce a Schr\"odinger cat like state of the bits
\cite{anders00,molmer99,sackett}. Throughout this Letter we apply a Pauli
spin notation for the  
description of the qubits. The qubit states $|0_l(1_l)\rangle$ are defined
as the $\sigma_{zl}=-1(1)$ eigenstates.

It will be useful to have a geometric picture of the contents of
Eq.~(\ref{milburn}): Each of the exponential terms on the left hand side
are displacement operators 
for the harmonic oscillator (conditioned on the internal eigenstates
of  operators $\hat{A}$ and $\hat{B}$), and the four terms displace the
system around the rectangular path in Fig.~1 (a). When a system is displaced
around a closed loop in phase space, the state vector  acquires
a geometric phase factor equal to the enclosed area \cite{anders00}. In
Fig.~1(a) the area is the product of the lengths of the sides, which due to
the operator character  
of these lengths becomes the product of two internal state operators, and
the resulting  phase factor is the operator on the right hand side of
Eq.~(\ref{milburn}).

\begin{figure}
\epsfig{width=8cm,file=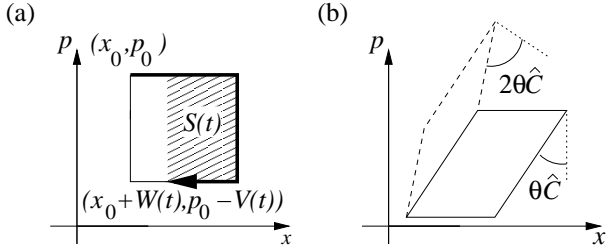}
\\
\caption[]{
Translations in $xp$-phase space of the oscillator during gate
operation: (a) In the evolution described by Eq.~(\ref{milburn}), the
oscillator is displaced by the amount $\lambda_2\hat{B}$ along the
x-axis, then by $-\lambda_1\hat{A}$ along the $p$-axis, etc., and when it
eventually ends up in the initial state, a geometric (and internal state
dependent) phase factor given by the enclosed area 
$\lambda_1\lambda_2\hat{A}\hat{B}$ multiplies the state vector of
the system. (b) By application of an interaction proportional to
$\hat{C}n$, the displacement along the $p$-direction  in part (a) of the
figure can be rotated into another direction given by the angle
$\theta \hat{C}$, and the area enclosed by the solid line becomes
$\lambda_1\lambda_2\hat{A}\hat{B}\cos(\theta \hat{C})$. To perform a
Grover search or C$^{n}$-NOT operation we need to enclose several 
parallelograms with angles which are multipla of $\hat{C}\theta$. An
effective way to  achieve this is to place all subsequent parallelograms so
that they share one side with the previous one, as
shown with the dashed curve for the second parallelogram. With this
construction, the multi-bit operation 
can be achieved by traversing only the outline of the combined figure.} 
\end{figure}

The trick contained in Eq.~(\ref{milburn}) suffices to produce two-bit gates
since the
operators $\hat{A}$ and $\hat{B}$ can be replaced by any pair of single
particle operators 
acting on particle one and two. 
The  C-NOT gate,
which is obtained by using $\hat{A}=(\sigma_{z1}+1)/2$,  
$\hat{B}=\sigma_{x2}$ and
$\lambda_1\lambda_2=\pi/2$, can be combined with single
particle operations to produce any unitary operation acting on all the
bits \cite{barenco}. This method in general involves several one and two
particle gates to produce multi-bit gates. For instance, in
Ref. \cite{barenco}  4 one-bit gates
and 3 two-bit gates were used to construct a three-bit gate which apart from
phase-factors is equivalent to the  C$^2$-NOT
or Toffoli gate. Experimentally each gate corresponds
to turning on a 
given Hamiltonian for a certain duration, and therefore each gate adds an
experimental  
complication and/or possibility of error. In this paper we pursue a
different strategy for implementing multi-bit gates. We will show that one
may extend the trick in Eq.~(\ref{milburn}) to produce higher order gates
directly. 

In \cite{anders00}, we discussed the application of a Hamiltonian
with continuously varying terms in $\hat{A}x$ and $\hat{B}p$,
and we showed, in particular, that harmonically varying coefficients
on the operators corresponding to bichromatic fields, can 
also be used to produce the operator products. To extend these result to
multi-bit gates we shall need a slightly more general interaction which may
be described by 
\begin{equation}
H(t)=v(t)\hat{A}x +  w(t)\hat{B}p + r(t)\hat{C} n,
\label{ham}
\end{equation}
where $\hat{A}$, $\hat{B}$ and $\hat{C}$ are commuting operators acting on
the internal states
of the atoms, $n$ is the number operator for the harmonic oscillator, and
$v$, $w$, and $r$ are arbitrary functions of time.
With this Hamiltonian the time
dependent Schr{\"o}dinger equation for the propagator $idU(t)/dt =H(t)U(t)$ 
has the solution 
\begin{equation}
 U={\rm e}^{-i\hat{S}(t)} {\rm e}^{-i n \hat{R}(t)} {\rm e}^{-i x
 \hat{V}(t)}  {\rm
 e}^{-i p \hat{W}(t)}, 
\label{propagator}
\end{equation}
with 
\begin{eqnarray}
\hat{R}(t)&=& \hat{C} \int_0^t r(t') dt'\nonumber \\
\hat{V}(t)&=& \int _0^t \hat{A} v(t') \cos(\hat{R}(t'))-\hat{B}
w(t')\sin(\hat{R} (t')) dt'\nonumber \\
\hat{W}(t)&=&  \int _0^t \hat{B} w(t') \cos(\hat{R}(t'))+\hat{A}v(t')
\sin(\hat{R} (t')) dt' \nonumber\\
\hat{S} (t)&=& -\int_0^t \hat{V}(t')[\hat{B} w(t') \cos(\hat{R}
(t'))\nonumber\\
&&\mbox{\hspace{3cm}}+\hat{A} v(t')\sin(\hat{R} (t'))]
dt'.  \label{functions}
\end{eqnarray}
It is straightforward to check the solution by taking the time
derivative of $U$ and using the Baker-Hausdorf relation to simplify the
result.   
In the $xp$-phase space the net action of this propagator is to perform
translations $(x,p)\rightarrow (x+\hat{W}(t),p-\hat{V}(t))$ followed by a
rotation by 
an angle $\hat{R}(t)$ around the origin. Since the
functions $\hat{V}$, $\hat{W}$, and $\hat{R}$ involve the internal state
operators $\hat{A}$,
$\hat{B}$, and $\hat{C}$ the translation and rotation is entangled with the
internal 
states of the bits. We now generalize 
the trick applied in Eq.~(\ref{milburn}) to ensure
that $\hat{V}$, $\hat{W}$, and $\hat{R}$ vanish after a certain time $\tau$,
such that the harmonic oscillator is returned to its initial state, and we are
left with an internal state evolution operator $\exp(-i\hat{S}(\tau))$,
where $\hat{S}(\tau)$ is the area enclosed by the trajectory in the phase
space.  
Note, that the expression for $\hat{S}(\tau)$ does not
involve operators referring to the harmonic
oscillator, so that the gate is insensitive to the initial state
of the oscillator. This gate can be applied with the oscillator in an unknown 
state, {\it e.g.}, in a thermal state. 

As a first concrete example of our procedure consider three bits which are
subject to the time independent Hamiltonian
\begin{equation}
H=\Omega {\left(\frac{\sigma_{z1}+\sigma_{z2}+1}{4\sqrt{K}} x -\sigma_{x3}
    {\left(n+\frac{1}{32K}\right)} \right)},
\label{htoffoli}
\end{equation}
where $K$ is an integer. After a duration $\tau=K2\pi/\Omega$ the
propagator (\ref{propagator}) reduces to
$\exp(-i\pi[({\sigma_{z1}+\sigma_{z2}+1})^2-1]\sigma _{x3}/16)=
\exp(-i\pi(\sigma_{z1}+1)
(\sigma_{z2}+1) \sigma_{x3}/8)$, which is exactly the Toffoli gate. 
(We used the fact that for a single particle Pauli operator $\sigma$,
$\sin(\theta \sigma)=\sin(\theta)\sigma$).
In the ion trap quantum computer the gate can be achieved by applying a
single pulse of suitably directed and detuned fields to the ions.

The three particle Toffoli-gate, can be constructed so easily because
the constant term in ${\sigma_{z1}+\sigma_{z2}+1}$ can be chosen so that
this operator squared yields the desired combination of internal state
operators apart  from a single particle rotation. This technique is not
directly applicable for more than three particles, and we have not been able
to devise a similar simple construction with only a single Hamiltonian in this
situation.  
Instead we shall produce gates by sequentially applying three different
Hamiltonians. To
make a C$^{n_c}$-NOT operation, where the first $n_c$ qubits control the
action of the $n_c+1^{st}$ qubit, we 
need a projection operator
which projects into the space where all the $n_c$ control bits are in the
$|1\rangle$ 
state. Such a projection operator can be expressed as a 
product of single particle projection operators $\prod_{l=1}^{n_c}
(\sigma_{zl}+1)/2$, and the  C$^{n_c}$-NOT operation may be expressed as
$\exp(-i\pi/2 \prod_{l=1}^{n_c} \frac{(\sigma_{zl}+1)}{2} \sigma_{x n_c+1})$.
The operators that are easy to make
in practice are {\it sums} of individual particle operators 
like $\hat{J}_z-J
= \sum_{l=1}^{n_c} \frac{(\sigma_{zl}-1)}{2}$.  
To turn the sum into a product, we observe that if and only if
all $n_c$ control qubits  
are in the $|1\rangle$ state, not only is the product $\prod_{l=1}^{n_c}
\frac{(\sigma_{zl}+1)}{2}$ equal to unity, also the sum 
$\hat{J}_z-J = \sum_{l=1}^{n_c} \frac{(\sigma_{zl}-1)}{2}$ vanishes.
We now use the Fourier transform
$\sum_{k=1}^{m} \cos(2\pi \frac{k}{m} N) = m\delta(N~{\rm mod}~m)$ 
which can also be applied to operators so that:
\begin{eqnarray}
\prod_{l}^{n_c}\frac{(\sigma_{zl}+1)}{2} = \frac{1}{n_c+1}\sum_{k=1}^{n_c+1} 
\cos{\left(\frac{2\pi k}{n_c+1}(\hat{J}_z-J) \right)}.
\label{magic}
\end{eqnarray}
The C$^{n_c}$-NOT gate is thus the product of $n_c+1$ 
terms $\exp(\frac{i\pi}{2(n_c+1)} \cos(\frac{2\pi
k}{n_c+1}(\hat{J}_z-J))\sigma_{x n_c+1})$ ($k=1,2, ... ,n_c+1$).

To implement a unitary operator which can be written in the form
$\exp(-i \mu \hat{A}\cos(\theta \hat{C}))$,
where $\hat{A}$ and $\hat{C}$ are internal state operators, we make explicit
use of the fact that we have an internal 
state operator appearing inside a trigonometric function in the expression for
$\hat{S}$ (\ref{functions}). Geometrically, 
we follow the construction of the parallelogram in Fig.~1 (b):
First, we apply a Hamiltonian proportional to $\hat{A}p$
which performs a translation along the $x$-axis. Then a Hamiltonian 
$H\sim \hat{C}n$ makes a rotation of the phase space by an angle
$\theta\hat{C}$:
 $\exp(i\theta \hat{C}n)x\exp(-i\theta \hat{C}n)=  
 \cos(\theta \hat{C})x + \sin(\theta \hat{C})p$, and we
perform a translation along the $p$-axis with $\hat{B}$ equal to the
identity, etc.  The enclosed
area is proportional to $\hat{A}\cos(\theta \hat{C})$ and the propagator
has the desired form $\exp(-i \mu \hat{A}\cos(\theta \hat{C}))$. By varying
the strength and duration  of the pulses one can control the parameters
$\theta$ and $\mu$, and  using 
$\hat{A}=\sigma_{x,n_c+1}$ and $\hat{C}=\hat{J}_z-J$
 the parallelogram results in the time evolution operator
$\exp(-i\mu
\cos(\theta (\hat{J}_z-J))\sigma_{x,n_c+1})$.

By using the operator
identity (\ref{magic}) we can devise a C$^{n_c}$-NOT gate by following
the outline of  $n_c+1$ such parallelograms, one after the other.
By rotating each parallelogram, so that the first linear displacement
is precisely the opposite of the last displacement of the previous
parallelogram, we can save half of the translations.
Note that the sum over $l$ implicit in 
the $\hat{J}_z$
term in Eq.~(\ref{magic}) just amounts to illuminating several qubits 
instead of a single qubit at a time. 

In 1997, Grover presented a search algorithm \cite{grover}
that identifies the single value $x_0$ that fulfills $f(x_0)=1$
for a function $f(x)$ provided, {\it e.g.}, by an oracle
(all other arguments lead to  vanishing values of the function).
If $x$ is an integer on the range between 0 and $N-1=2^n-1$, the search
algorithm is able to find $x_0$ after on the order of $\sqrt{N}$ 
evaluations of the function. Grover's algorithm has been 
demonstrated on NMR few qubit systems \cite{nmr_grover}. In the following
we show how our proposal can be used to implement the search algorithm. 

The quantum algorithm first prepares an initial trial state vector 
populating all 
basis states with equal probability. To implement a full Grover search the
function $f(x)$ has to be a non-trivial function which is implemented by
the quantum computer, but for demonstrational purposes, 
the function $f(x)$ can be encoded by letting the state of the register 
undergo a transformation where the amplitude of the $x_0$ component 
changes sign and all other amplitudes are left unchanged. This step 
can be implemented by writing $x_0$ in binary form, $b_0b_1b_2... b_{n-1}$, 
and by applying the unitary operator
\begin{eqnarray}
U_f = \exp{\left( i\pi \prod_{l=0}^{n-1} {\left(\frac{\sigma_{zl}+2b_l-1}{2}
  \right)}   \right)}.
\label{function}
\end{eqnarray}
Below we show how this time evolution may be implemented with our
procedure. Note that the corresponding effective 
Hamiltonian vanishes when
applied to any state where the qubit value (eigenvalue of $\sigma_{zl}$)
does not coincide with $2b_l-1$, {\it i.e.}, the state must be 
the exact representation of $x_0$ to acquire the sign change.

The crucial step in Grover's algorithm is an `inversion about
the mean', where the amplitude with the sign changed will grow in comparison
with the other amplitudes. In the $n-$qubit computer with $N=2^n$
amplitudes $c_x$, the operation can be written $c_x \rightarrow
\frac{1}{N}\sum_{x'} c_{x'} - (c_x-\frac{1}{N}\sum_{x'} c_{x'})$.
The sum of all
amplitudes of the state vector $|\psi\rangle$ can be obtained as any 
component in the vector $M |\psi\rangle$, where $M$ is the $N \times N$
matrix with unit elements in all positions.
The inversion about the mean is therefore given by the unitary matrix
\cite{grover}
\begin{eqnarray}
U_G = \frac{2}{N}M - I,
\label{ug}
\end{eqnarray}
where $I$ is the $N\times N$ identity matrix.

A straightforward calculation shows that the $M$ matrix fulfill 
$(sM)^k=s^kN^{k-1}M$, and hence we have the  exponential
\begin{eqnarray}
\exp(sM) = I + \sum_{k=1}^{\infty} \frac{(sM)^k}{k!} = I +
\frac{1}{N}(e^{sN}-1)M.
\end{eqnarray}
Thus, by choosing $sN=i\pi$, we get $\exp(sM) =I-\frac{2}{N}M$,
which apart from an irrelevant global phase yields precisely
the inversion about the mean.

In the standard binary basis, the matrix $M$
couples all states to any other state, and it can be written as the
tensor product $\Pi_{l=0}^{n-1} (\sigma_{x l}+1)$, where the single
qubit operators  $\sigma_{xl}+1$ are $2\times 2$ matrices with
unit elements in all positions. 
The inversion about the mean is therefore produced directly by the 
action of the following multi-particle operator 
\begin{eqnarray}
U_G = \exp{\left(i\pi \prod_{l=0}^{n-1}
    {\left(\frac{\sigma_{xl}+1}{2}\right)}  \right)},
\label{grover}
\end{eqnarray}
where we used $N=2^n$.

Both $U_f$ and $U_G$ can be implemented effectively using (\ref{magic}).
To implement the function (\ref{function}), 
it is easiest to first invert all the bits,
which have the value zero in $x_0$, so that $U_f$ on that state
should encode only unit bit values, {\it i.e.}, $U_f$ is precisely the
exponential of the projection
operator in the left hand side of Eq. (\ref{magic}). Following the outline
of the parallelogram in 
Fig. 1 (b) with $\hat{A}$ and $\hat{B}$ equal to the identity and
$\hat{C}=\sum_{l=0}^{n-1} \frac{(\sigma_{zl}-1)}{2}$ we obtain the
exponential of one of
the terms in the sum on the right hand side, and by combining $n+1$ such
terms one can construct the full sum.
After application of this simple $U_f$,
the qubits encoding the value zero should be flipped back again. 
All qubits should then have their $\sigma_x$ components rotated into
the $z$-direction, to use again the operation in (\ref{magic}) to implement 
$U_G$, which is the same operator, defined for the $x$-components of the
spins. The whole algorithm only requires individual access for the 
single qubit spin flips, encoding $x_0$, and for the final readout.
An easy demonstration experiment where $x_0=1111 ... 1$ can thus be performed
without individual access at all (one only needs to verify that  the number
of excited qubits at the end equals the total number of qubits).

In summary, we have presented a technique to produce multi-bit gates
in quantum computers where all qubits are coupled to a joint
harmonic oscillator degree of freedom. We have derived general
expressions, and we have exemplified the method by an analysis of
the Grover search and  the C$^n$-NOT gate, which
appears frequently, {\it e.g.}, in error  correcting codes \cite{Steane}.
A recent preprint \cite{concurrent} has addressed the achievements of
so-called `concurrent 
quantum computing', in which access to multi-bit interaction
Hamiltonians of the form $\Pi_l \sigma_{zl}$ is assumed. That paper
presents ideas for Grover's and Shor's algorithm, without suggesting
a practical means to implement the interaction. Our procedure
provides a proposal for such implementation: since we can write $\exp (-i
\mu \prod_l \sigma_{zl})=\exp[-i\mu \cos(\pi \sum_l (\sigma_{zl}-1)/2 )]$, 
a single parallelogram as in Fig.~1 (b) suffices to produce this operator. 

It is known how to make C$^2$-NOT and C$^3$-NOT gates by means of 
one- and two-bit gates, but it is difficult to make a theoretical
comparison of these implementations with our proposal, since we build 
up the desired \mbox{one-,} \mbox{two-,} and multi-bit interactions
continuously in time.
From a practical
perspective, however, our scheme should be really advantageous.
The essential operation in the Grover search (\ref{grover}) is
implemented without access to the  individual qubits and, {\it e.g.}, in
the ion trap 
it is much easier to implement the Hamiltonian $H=\sum_l
(\sigma_{zl}-1)n$ than just a single term
$H=(\sigma_{zl}-1)n$ in that sum. 
In addition, it is an experimental advantage to apply as
few control Hamiltonians as possible, since imprecision in timing
accumulates if many operations are needed.

A feature of our proposal worth emphasizing is that all operations
are expressed as unitary gates acting on the qubit degrees of freedom.
The oscillator is certainly important, and only at the end of the gates,
do the qubits  actually decouple from the oscillator. One consequence is that
the initial state of the oscillator does not have to be specified.
It can be in the ground state, an excited state, or even in an incoherent
mixture of states, possibly entangled with environmental degrees of
freedom, as long as this entanglement does not evolve during gate
operation. 

This work was supported by the Danish National
Research Council, Thomas B. Thriges Center
for Kvanteinformatik, and by the Information Society
Technologies programme IST-1999-11053, EQUIP, action
line 6-2-1.


\end{document}